\documentstyle[aps,prl,preprint]{revtex}
\include{psfig}
\begin{document}
\draft \title{\bf Supercooled Water and the Kinetic Glass Transition II:
Collective Dynamics}

\author{Francesco Sciortino$^{\dagger}$, Linda Fabbian$^{\dagger}$,
 Sow-Hsin Chen $^{\dagger\dagger}$ and Piero Tartaglia$^{\dagger}$}

\address{$^{\dagger}$ Dipartimento di Fisica and Istituto Nazionale
 per la Fisica della Materia, Universit\'a di Roma {\it La Sapienza},
 P.le Aldo Moro 2, I-00185, Roma, Italy}

\address{{$^{\dagger\dagger}$} Department of Nuclear Engineering,
Massachusetts Institute of Technology, Cambridge, MA 02139\\}

\date{\today} \maketitle

\begin{abstract}
In this article we study in detail the $Q$-vector dependence
of the collective dynamics in simulated deeply supercooled 
SPC/E water. The evolution of the system has been followed for
250 $ns$ at low $T$, allowing a clear identification of a
two step relaxation process. 
We present evidence in
favor of the use of the mode coupling theory for
supercooled liquid as  framework for the 
description of the slow $\alpha$-relaxation 
dynamics in SPC/E water, 
notwithstanding the fact that
the cage formation in this system is controlled by the
formation of an open network of hydrogen bonds as opposed to
packing constraints, as in the case of simple liquids.

\end{abstract}


\section{Introduction}
\label{intro}
The slow dynamics ($\alpha$-relaxation) in supercooled
molecular liquids and the glass-transition 
are two related topics which have received
particular attention from the scientific community in the last years
\cite{review,review-glass}.
Significant progress have been made by a sinergetic approach
based on theoretical, experimental and more recently on numerical work.
Nowadays, numerical simulations with realistic potentials describing the
evolution of molecular system composed by thousand atoms, 
for time interval longer than $100~ns$, are becoming feasible, 
allowing a closer check of the theoretical predictions and
bridging the gap between experiments and theory.
Such long simulations, although are suffering on the
deficiency of the potential used compared to the real systems 
which they aim to simulate, offer a ideal reference for 
comparing with the novel theoretical predictions concerning 
the slow dynamics above the glass transition.

In the case where the studied system is water, the interest 
in interpreting  the molecular dynamics in term
of glass-transition concepts goes beyond the comparison between
the simulated system and the theoretical predictions. Indeed,
if the experimentally observed non-Arrhenius increase of
the transport coefficients on supercooling\cite{AngRev,lude}  
can be explained within the same framework of simple
supercooled liquids\cite{Germans,johari},
the presence of 
hidden  thermodynamics anomalies at ambient pressure\cite{Speedy,SpeedyAngell} 
does not 
have to be invoked\cite{sri96,fsvarenna}.
In the first paper of this series\cite{sgtc}
we have tested Mode Coupling Theory (MCT)\cite{lastgotze} 
predictions for the correlation
functions of single particle dynamics in water with corresponding
quantities calculated from 
Molecular Dynamics (MD) simulations,
carried out for sufficiently long time 
($20$ to $50~ns$ ) as to allow the slow dynamics
to be observed.  We tried to assess to what extent the MCT,
which has been shown to describe simple liquids\cite{KobAnd,kob}, is 
applicable also to the description of the single particle dynamics of 
(simulated) supercooled water, an hydrogen bonded liquid with strong
non-isotropic interactions among molecules. 
As a result we found that the 
center of mass tagged particle dynamics can be qualitatively
interpreted in terms of MCT. This result stimulated us to
make the comparison more stringent and to extend it to the
collective center of mass dynamics.
With one further year
of computer time on three workstations working full-time, 
the limit of $250~ns$ of continuous simulation time 
has been reached, allowing the
calculation of the collective properties, for which the noise level
can not be reduced averaging 
over different molecules in the simulated system.

Theoretical work based on the MCT has mainly focused on
simple atomic liquids and on molecular liquids 
with spherical-symmetric interactions.
Only recently
the theory has been extended to treat the case of molecular
systems with non-isotropic interaction potentials  
opening the way for a full and detailed quantitative comparison between
theory and simulation/experiments for molecular systems\cite{walstrom}. 
Unfortunately,
preliminary results at the ideal MCT level are only available for
dipolar hard spheres\cite{schilling-andalo,kob-dipolar}, i.e. molecules
with a simpler geometry than the water molecule geometry. 
Such results\cite{schilling-andalo} 
strongly supports the validity
of general predictions based on the MCT for simple liquids.

In this article we study in detail the $Q$-vector dependence
of the collective dynamics in SPC/E water and present evidence in
favor of a MCT description of the slow $\alpha$-relaxation dynamics.
The comparison is performed on a two level scale:
(i) At qualitative level,
we compare the MCT prediction for simple liquids
(exemplified by the hard-sphere liquid within the Percus-Yevick
approximation) with the center of mass dynamics in the case of 
supercooled SPC/E water. 
Although the comparison is by default qualitative, we think that the
analogy in the $Q$ dependence of all relevant parameters is 
particularly significant.
(ii) At quantitative level, we
compare the SPC/E slow dynamics with the dynamics predicted by MCT
in the late $\beta$-region; moreover we evaluate
the von Schweidler exponents $b$ and $\gamma$ governing the relaxation
process (Sec.\ref{sec:theory}) and we verify that they are related 
as MCT predicts. 

In both cases the agreement is 
striking and strongly supports the validity of MCT as the correct framework for
interpreting the slow collective dynamics in simulated 
supercooled water.

\section{MCT}
\label{sec:theory}

Several review papers present in detail the theory of mode coupling
for supercooled liquids, in particular in its {\it ideal} 
formulation. A review on the predictions of the theory
can be found in Ref.\cite{sgtc}.
In this section we briefly report the main results which 
are relevant for the reading of the article.
 
MCT aims at describing the slow dynamics in supercooled glass forming 
simple liquids or in molecular liquids with spherical-symmetric 
interactions between the molecules. 
It provides a description of the time evolution of density and current
correlation functions (correlators in the following) in the time region where
structural relaxation becomes the process which entirely controls the dynamics,
{\it i.e.} for times longer than the microscopic characteristic time. 
MCT describes the slowing down of the structural relaxation, which is
typical of supercooled liquids, including memory effects in 
a memory function  which depends only on statical quantities 
(number density and structure factor $S(Q)$). 

In the {\it ideal} formulation the loss of correlation is
ascribed completely to interaction between fluctuations of 
density pairs; all other channels for the decay of correlation, as
for example the momentum modes, are completely neglected. 
The ideal MCT predicts a sharp liquid to glass transition, located at a finite 
critical temperature $T_c$, associated with a power-law
divergence of the correlation time. 
$T_c$ is a purely kinetic transition temperature which describes 
the freezing of the molecules inside the cages and it does not deal with
any thermodynamical anomalous behavior.  
The {\it ideal}-MCT can be considered as a first order approximation 
in a more complex scheme \cite{gotze-book}; the 
kinetic transition is an artifact of the approximations
involved and it disappears in
the {\it extended} version of MCT, where couplings with
particle momenta are taken into account. The power-law
increase of the characteristic times predicted by the
ideal MCT for $T>T_c$ crosses to a different
(unknown) relation, to describe the complex activated 
dynamical processes below $T_c$. The concept of $T_c$ 
retains the meaning of cross-over temperature between two
different  dynamical behaviors and the prediction of
the ideal MCT can be used to interpret the slow dynamics
above $T_c$, when hopping effects are negligible. 

The ideal MCT predicts the existence of a two step relaxation
scenario for the slow dynamics in supercooled simple liquids
at temperatures close to $T_c$. 
The decay of the correlators, {\it i.e.} 
the loss of memory of the initial configuration,
occurs in a two-step process\cite{review-glass,Got}: first the 
normalized correlator $\phi(t)$ approaches a plateau value $\phi_{EA}$ 
which is temperature-independent 
(non-ergodicity or Edwards-Anderson parameter); then  
$\phi(t)$ decays from $\phi_{EA}$ to zero 
($\alpha$-relaxation region). 
The short times region is strongly affected 
by the microscopic dynamics, which instead completely disappears in the long
time region ($\alpha$-relaxation).
In the region close to the plateau ($\beta$-relaxation region)  
$\phi(t)$  can be asymptotically expanded near the value $\phi_{EA}$:

\begin{equation}
\phi(t) = \phi_{EA} +  h \sqrt {\sigma}  g(t/t_\sigma)
\label{eq:beta}
\end{equation}

where $t_\sigma$ is the time scale which characterize the $\beta$-region
and  $\sigma$ is the 
separation from the critical point (at fixed number density
$\sigma=(T-T_c)/T_c$).
The function $ g(t/t_\sigma) $ (also called $\beta$-correlator)
is solution of the asymptotic equation

\begin{equation}
-1 + \lambda \tilde{z} LT [g^2(\tilde{t})] (\tilde{z})+
[\tilde{z}g(\tilde{z})]^2=0
\label{eq:equation}
\end{equation}

where $\tilde{t} = t/t_\sigma$ is the scaled time, $LT$ means the 
Laplace transform and $\tilde{z}=z t_\sigma$ is the Laplace 
conjugate of $\tilde{t}$; $\lambda$ is the exponent 
parameter\cite{gotze-book}. 

It's easy to show that Eq. \ref{eq:equation} has a scale invariant solution;
in the early $\beta$-region, {\it i.e.} for $t << t_\sigma$,
$g(\tilde{t})$ has, at the leading order, the fractal behavior:

\begin{equation}
g(\tilde{t})=\tilde{t}^{-a} +O(\tilde{t}^{a})
\end{equation}

In the late $\beta$-region ($t>>t_\sigma$) the predicted scaling law
is

\begin{equation}
g(\tilde{t})=- \tilde{t}^{b} +O(\tilde{t}^{-b})
\end{equation}
 
The two scaling exponent $a$ and $b$ are not independent; they are both
determined by the exponent parameter $\lambda$ through the relation:
\begin{equation}
\lambda = {{\Gamma(1-a)^2} \over {\Gamma(1-2a)}}=
{{\Gamma(1+b)^2} \over {\Gamma(1+2b)}}
\label{eq:lambdaab}
\end{equation}

where $\Gamma$ is the Euler's gamma function.
The range of validity of these expansions for $g(\tilde{t})$  is
limited at small $t$ by the microscopic dynamics time and at
large $t$ by the $\alpha$-relaxation time. 

In the $\alpha$-relaxation region another time scale appears, $\tau$;
in terms of the rescaled time $\hat{t}=t/\tau$ the behavior of the correlator
near the plateau is described in the leading order by a power law dominated 
by the exponent $b$.
Including the next to leading order corrections the departure from $\phi_{EA}$
in the early $\alpha$-relaxation region is given by
\begin{equation}
\phi(t) = \phi_{EA} - h_{(1)} \hat t^b + h_{(2)} \hat t^{2b} + O( t^{3b})
\label{eq:von}
\end{equation}
The amplitudes $h_{(1)}$ and $h_{(2)}$ strongly depend on the physical 
features of the studied liquid and they  have been 
explicitly calculated for a few models \cite{fuchs,mayer}.
The $\alpha$-relaxation time scale is a temperature dependent parameter
which scales as the inverse of diffusivity:

\begin{equation}
\tau(T) \sim |T-T_c|^{-\gamma}
\label{eq:gamma}
\end{equation}
with 

\begin{equation}
\gamma =1/2a+1/2b
\label{eq:abgamma}
\end{equation}

The three scaling exponents $a$, $b$ and
$\gamma$ are not universal since they depend on the static 
structure factor.
In the $\alpha$-relaxation region the
correlators obey the so-called time-temperature superposition
principle, i.e. it is possible to scale
the same correlator evaluated at different $T$ on a single
master curve, i.e.
\begin{equation}
\phi(t)=\phi(t/\tau(T))
\label{eq:master}
\end{equation}

For times much longer than $\tau$ the mode-coupling dynamical equations 
can not be solved either analytically nor asymptotically.
From numerical solutions developed for simple potentials 
it has been shown that a  Kohlrausch-Williams-Watts (KKW)  stretched
exponential form, i.e.
\begin{equation}
\phi(t) = A_{K} exp[-({t \over \tau_{K}})^{\beta_{K}}]
\label{eq:K}
\end{equation} 
can be used to empirically fit the last part of the $\phi(t)$ decay.

If we specialize the previous equations to the case in which $\phi$ is the 
intermediate scattering function
$F(Q,t)$, the self correlation function $F_{self}(Q,t)$ or the 
transverse current correlation $J_t(Q,t)$, 
the non-ergodicity factor $\phi_{EA}$,
the critical amplitudes $h$, $h_{(1)}$, $h_{(2)}$, and the fitting
parameters $\tau_K$, $\beta_K$ becomes $Q$-dependent quantities. 
Therefore the comparison between
theory and experiments can be extended from the time dependence
of $\phi$ and the temperature dependence of the correlation times to the
$Q$-dependence of the above mentioned parameter.

The MCT set of coupled  integro-differential  equations  can be 
numerically solved once the static structure factor $S(Q)$ is known.
The problem has been solved for many different potentials:
hard spheres \cite{fuchs,barrat-hard}, soft spheres \cite{barrat-soft}, 
Lennard-Jones \cite{beng}, binary mixtures of soft spheres \cite{barrat-binary} 
and of Lennard-Jones \cite{kob-binary}. 
For all these potentials the 
$Q$-dependence of the quantities appearing in the previous equations
as well as the fitting parameters ($\tau_{K}$ and $\beta_{K}$) 
in Eq. \ref{eq:K}
shows pronounced oscillations in phase with $S(Q)$.

Moreover it has been shown theoretically\cite{review-glass,Got,gotze-book} 
that the large $Q$ 
limit of the fitting parameters $\beta_K$ coincides with the scaling 
exponent $b$. 

\section{simulation data}

The MD data analyzed in this article are a series of  250
ns long 
trajectories for a system of $N$=216 water molecules interacting with the SPC/E
potential with periodic boundary conditions. 
SPC/E, one of the  most common water-water potential, is 
a pair-wise additive potential designed to mimic 
the interaction between {\it rigid} water molecules via electrostatic and
Lennard-Jones interactions. 
The simulation technique
has been discussed in detail in Ref.\cite{sgtc} and it is not repeated
here.  The only difference between the data analyzed in this article
and the data analyzed in Ref. \cite{sgtc} is the length of the
simulation which now cover a time interval more than 5 times longer.
We study seven different temperatures, ranging from $T=207 K$ up to $285 K$.

\section{analysis}

The basic quantity in the study of the center of mass  
collective dynamics in a liquid is the coherent 
intermediate scattering function, defined as

\begin{equation}
F(\vec Q,t)={1 \over S(Q)} \sum_{j,k=1}^{N} e^{-i \vec Q [\vec r_k(t)-\vec r_j(0)]} 
\end{equation}

where $N$ is the number of molecules in the system, $\vec Q$ is the wavevector
and $\vec r_i$ is the position of the center of mass of molecule $i$.
$F(\vec Q,t)$ is the autocorrelation function of the
space-Fourier transform of the density, thus giving 
information on the decay of density fluctuations at wavevector $Q$. 

The $T$ and $Q$ dependence of $F(Q,t)$ are shown in
Fig.\ref{fig:fqt-T} and \ref{fig:fqt-Q}. We show the $T$ dependence at
two different $Q$ vectors, namely (i) the $Q=18 nm^{-1}$ vector, 
corresponding to the
position of the first pre-peak in the structure factor, $Q_{FSDP}$ 
(the analogous of the so-called
first sharp diffraction peak in silica); (ii) the $Q=30 nm^{-1}$ 
vector corresponding to  the most intense peak in the structure factor, 
$Q_{SQMAX}$.
In agreement with the general MCT predictions,
all correlators show a two step decay process. For times smaller than $4 
ps$ we observe an oscillatory 
decay process, from one to the non-ergodicity parameter. This time window
contains
all information about the microscopic dynamics and about the
collective sound propagation in the $Q$ range  compatible 
with the simulation box size, a topic which have been 
studied in detail in several previous papers, independently from
the MCT modelization and which will not be discussed in this
article. We refer the interest reader to Refs.\cite{sound}, where the
frequency range above $10 cm^{-1} \sim 2\pi/4~~~ps^{-1}$ 
has been studied in detail.
The second decay process, from  the non-ergodicity parameter to zero,
has a strong $T$ and $Q$ dependence. This monotonic decay
(the $\alpha$-relaxation)
becomes slower and slower on decreasing the temperature.
At the lowest simulated $T$, the $\alpha$-relaxation is clearly separated 
from the microscopic time window and
appears in a time region which was never explored before 
due to the long cpu time required to perform equilibrated 
molecular dynamics simulations. 

To quantify the $Q$ and $T$ dependence of the $\alpha$-relaxation
we fit all correlators for times longer than 7 ps 
(to avoid the interference of the oscillatory sound modes)
with a stretched exponential function  (Eq.\ref{eq:K}). 
For $T$ higher that 230 K, the $\alpha$ relaxation time
is smaller than 20 ps and the $\alpha$ relaxation process, being 
superimposed to the collective sound modes, can be fitted 
with different parameters. Instead, $F(Q,t)$ of the 
four lowest studied temperatures can be  fitted unambiguously.
The $Q$ and $T$ dependence of $A_{K}$, 
$\tau_{K}$ and $\beta_{K}$, at the lowest four studied temperatures, 
are shown in Fig.\ref{fig:fit}, together with $S(Q)$. 
We note the presence of oscillations in all fitting 
parameters highly correlated with the oscillations in $S(Q)$.

To stress the similarity of our center of mass results
with the prediction of MCT for spherical molecules we 
show  in Fig.\ref{fig:fit-py} 
~~$A_{K}$,$\tau_{K}$ and $\beta_{K}$ evaluated from the
numerical solution of the ideal MCT coupled 
integro-differential equations 
for the Percus-Yevick solution of the hard-sphere potential
The MCT equations for the $\alpha$-relaxation region\cite{review-glass} 
have been solved
using the same technique as in Ref.\cite{fuchs}. We note that
both in the theoretical hard-sphere result and in the case of SPC/E
center of mass dynamics, $\tau_{K}$ and $\beta$ oscillate in phase with
the structure factor. 
MCT also predicts that the limit of $\beta_{K}$
at large $Q$ coincides with the value of the
exponent $b$ in Eq.\ref{eq:von}. 
From Fig.\ref{fig:fit} we note that
at large $Q$, $\beta_{K}$ tends to the value $0.5$, the same value
estimated previously on the basis of the analysis of the
tagged particle motion for SPC/E water.

According to MCT, close to $T_c$,
the $T$ dependence of  $\tau_{K}(Q)$ at any fixed $Q$ 
should follow Eq.\ref{eq:gamma}. For the case $b=0.5$, MCT predicts
for $\gamma$ a value of $2.7$. To compare with our data, we show
in Fig. \ref{fig:tau-T}  
$\tau_{K}(Q)/\tau_{K}(Q_{FSDP})$, i.e. imposing the
equivalence of the scaled time at $Q=Q_{FSDP}$.
In a large $Q$ range, i.e. $Q < 40 nm^{-1}$ all curves 
are nicely superimposed. We note that the longest relaxation time
coincides with the position of the first sharp diffraction peak,
as opposed to $Q_{SQMAX}$. Thus, the 
medium range order characteristic of network forming liquids\cite{elliot}, 
is the
most stable structure in the system. The enhanced stability at $Q=Q_{FSDP}$
can be predicted independently from $MCT$ by calculating
$S(Q)/Q^2$, i.e. on the basis of the so-called de Gennes narrowing.
The inset in Fig.\ref{fig:tau-T} 
shows that the $T$ dependance of the scaling factor $\tau_{K}(Q_{FSDP})$ 
is compatible
with a power-law with exponent $\gamma =2.7$ and $T_c=202 K$.
We stress that the same value for $\gamma$ was found for the
$T$ dependence of the self diffusion constant, but with 
$T_c=199 K$. The 2\% difference in $T_c$ is within the numerical error.
The time-temperature prediction of MCT, i.e. the fact that all correlators
have the same $T$-dependence of their relaxation time is consistent with the
equal value of $\gamma$ found for the self and the collective dynamics
in SPC/E water.
As we discussed before in Ref.\cite{sgtc}, 
the value of $\gamma$ calculated for SPC/E is
different from the one obtained by fitting the experimental data for
viscosity, diffusivity or reorientational 
NMR times --- which run from 1.6 to 2.4 on increasing the pressure. 
An extension of the present study at different isobars would be very valuable
for clarifying the ability of the SPC/E potential to describe the dynamics of
real water.

We now turn to the SPC/E dynamics in the region 
where the value of the correlators is not very
different from the non ergodicity factor, and where according to MCT 
the evolution of the
correlators is controlled by Eq.\ref{eq:master}. 
According to MCT, curves at different $T$ but at the same $Q$ values 
fall on the same master curve. Following the procedure suggested 
in Ref. \cite{KobAnd} we show in Fig. \ref{fig:master} 
$F(Q,t)$ vs. $t/\tau_{\alpha}(T)$ where $\tau_{\alpha}$ is
defined by $F(Q,\tau_{\alpha})=1/e$, i.e. it is the time at which the
correlation function has decayed to the $1/e$ value.
From Fig. \ref{fig:master} it is clearly seen that, at all $Q$,
curves for all $T$ tends to sit on the same master curve, even the
high $T$ systems for which the $\alpha$-region is hard to detect.
To make the comparison with $MCT$  more stringent, we compare
the master curve designed by the envelope of the different samples
with the universal master curve for the $\beta$-region, i.e. the
solution of Eq. \ref{eq:equation}. As discussed above, the time dependence
of $g(t)$ is controlled  only by the value of $b$, via
Eq.\ref{eq:lambdaab}. Thus we find that
the same $b$ value of $0.5$, which was estimated from the
study of self-dynamics in SPC/E water is able to rationalize both the
limit value at large $Q$ of $\beta_{K}$  as well as the
time behavior of all correlation function in the $\beta$-region.
Unfortunately, no information can be obtained from our data concerning the
exponent $a$, due to the superposition of the intense oscillations
related to sound modes to the critical decay. 

As predicted by the theory, the range of validity 
of  $g(t)$ is different for different $Q$ vectors,
and appears to be larger for $Q$ close to $Q_{FSDP}$ and smaller for
$Q_{SQMIN}=37 nm^{-1}$. 
The $Q$ dependence of the correction to the
master curves for large times (see Eq. \ref{eq:von}) 
are also predicted by MCT\cite{mayer} and 
can be compared with the center of mass collective dynamics in
SPC/E water. To this aim we fit $F(Q,t)$ at different $Q$ at
the lowest simulated temperature with the expression in Eq.
\ref{eq:von} imposing $b=0.5$, for all $Q$ values.
All correlator are fitted with the same $b$ value.
The quality of the fit is shown in Fig.\ref{fig:fitdue}, confirming that
at all $Q$ a rather good representation of the decay of correlation in the
early $\alpha$-region can be obtained in terms of Eq.\ref{eq:von}.
Fig. \ref{fig:fitdue} shows also that the contribution of
the next-to-leading order correction is not at all negligible
at some $Q$ vectors\cite{fuchs,mayer}.  
The result of the fit in the entire $Q$ range  are shown in 
in Fig.\ref{fig:h1h2}. Fig.\ref{fig:h1h2-py} shows the same quantities
calculated theoretically 
for the hard-sphere case (PY approximation). As in the
previous qualitative comparison, both $h_{(1)}(Q)$ and $h_{(2)}(Q)$
have oscillations in phase with $S(Q)$. 
Results in  Fig.\ref{fig:h1h2} clearly highlight the need of performing an
analysis in term of leading and next to leading expansion to detect the
correct exponent $b$. 
In any case, as already discussed in Ref. \cite{mayer},
$F(Q,t)$ for large $Q$ values
(large meaning larger than $Q_{SQMAX}$) is
not a good candidate for the identification of a von-Schweidler law.

\section{conclusions}

In this article we have presented evidence in favor of a MCT description 
of the slow collective 
dynamics in deeply supercooled simulated SPC/E water.
The presented data concerning the collective longitudinal 
dynamics, together with the data in Ref.\cite{sgtc} 
concerning the self motion, offer 
a complete picture of the dynamics of SPC/E water under deep
supercooling conditions. 
The evolution of the system has been
studied for more than 250 ns at low $T$, allowing for the first time
a clear identification of the two-step processes present in the
correlators decay.
The second decay process, clearly identified with 
the $\alpha$-relaxation decay, is characterized by a highly non 
exponential behavior and by an apparent divergence of the
characteristic time  at a temperature around 200 K, i.e. about 50 K
below the temperature of maximum density for SPC/E. An apparent divergence
of the transport coefficient in real water, again at about 50 K below the 
temperature of maximum density, is also observed 
experimentally\cite{A82,deben}.

The agreement between the data and MCT is striking both at 
qualitative and at quantitative level. 
At a qualitative level we have found that for both self and 
collective properties.
(i) All correlators decay with a two step process which spreads
over several time decades.
(ii) The $\alpha$-relaxation decay has an initial
power-law behavior on leaving the plateau, whose range of validity is 
$Q$-dependent; the exponent in Eq. \ref{eq:master} is  $b=0.5$; 
unfortunately an estimate of the
exponent $a$ is not feasible, due to the overlap in time with the
sound modes.
(iii) The long times part of the decay can be well fitted 
by a stretched exponential function.
(iv) The $Q$-dependence of the parameters in the
stretched exponential ({\it i.e.} amplitude, time and exponent)
oscillates in phase with the static structure
factor, in close analogy with the MCT predictions for simple
liquids.
(v) The $Q$-dependence of the von Schweidler amplitude and of the
next-to-leading corrections are similar to the one predicted
by MCT for a hard sphere system in shape and order of magnitude.

At quantitative level we have found that 
(i) The stretching exponent $\beta$ at large $Q$ tends to
0.5, the same value as $b$ as predicted.
(ii) All characteristic times,
both self and collective, satisfy the scaling law (Eq.\ref{eq:gamma})
in a large $Q$ range; moreover, the value of the
exponent $\gamma$ is  consistent with the value expected
from the knowledge of the $b$ exponent (Eq. \ref{eq:abgamma})
and it is independent from the correlator type.
(iii) The correlator time dependence for values close to the non-ergodicity
parameter is well described by the $\beta$-correlator 
(Eq.\ref{eq:equation}).

The analysis presented here and in Ref. \cite{sgtc} 
represents an important step towards the understanding of slow 
structural relaxation in complex glass forming liquids,
independently from the ability of SPC/E of mimicking real water.
From the theoretical point of
view, SPC/E is a non trivial molecular system, with molecules
interacting via a strong anisotropic pair-wise additive interaction
potential.  Structural arrest in this system is not driven by packing
constraints but by the formation of a strong network of tetrahedrally
coordinated water molecules. Molecules in the cage are hindered in
their translational and rotational motion by the presence of strong
hydrogen bonds (energetic cages). 
The fact that MCT, which has been developed for simple liquids,
succeeds well in describing the low $T$ center of mass 
dynamics in such a complicated molecular
system, strongly supports the existence of a universality in the
self and collective behavior of liquids under deep supercooling.
Such universality has begun to emerge in the first applications 
to non-spherical molecules of the recently developed extension of
the ideal MCT approach to treat angular correlators\cite{schilling};
this opens the way for
an even more quantitative description of the SPC/E dynamics 
in a MCT framework. 
The series of simulation here analyzed can become a clean reference
system to check the novel theoretical developements.

The analysis of SPC/E water presented here and in Ref.\cite{sgtc} 
is also important in respect to the thermodynamic behavior of 
simulated water. We refer the interested reader to the recent
book by Debenedetti \cite{deben} and to 
Ref. \cite{fsvarenna,sri96,PSES92,poole,SEPS97}. 
In this context, if SPC/E could be considered a 
sufficiently approximated modelling of real water, 
the data presented in this article would suggest that
(if hopping processes did not intervene)
at low pressure water would undergo a kinetic glass
transition about 50 degrees below the temperature of maximum density, 
suggesting
an interpretation of the so-called Angell 
temperature\cite{Speedy,SpeedyAngell,AngRev} as the critical
temperature of MCT\cite{GSTC,sgtc}.  In this regard, 
the apparent power-law increase
of transport coefficient in liquid water on supercooling is traced to
the formation of cages and to the associated slow dynamics resulting
from the presence of long living energetic cages.  
In other words the divergence
of transport coefficients does not need to rely on a thermodynamical
instability, either connected to the re-entrance of the gas-liquid
spinodal or to the presence of a critical point at high pressure and
low temperature\cite{PSES92,SEPS97}
If SPC/E can represent the thermodynamic behavior of water sufficiently, 
we would conclude that at low
pressure there is a continuous path connecting the liquid
state to the low density amorphous ice.

\section{Acknowledgments}
We thank F. Thiery for providing us the data shown in Fig.\ref{fig:fit-py}
and M.E. Camarda for support. This work is supported by grants from
GNSM/CNR and INFM/MURST.

\begin{figure}
\caption{$F(Q_{FSDP},t)$  (A) and $F_{SQMAX},t)$ (B) vs time.   
$\bigcirc~T~=~207.~K$, $\sqcup$\llap{$\sqcap$}~$ 
T~=~210.~K$, $\diamondsuit$~$T~=~213.2~K$,
$\triangle$~$T~=~225.0~K$, $\triangleleft$~$T=238.2 K$,
$\bigtriangledown$~$T=258. K$, $\triangleright$~$T=285. K$
Solid lines are fits with the KKW law (Eq.\protect\ref{eq:K})
for times longer than 7 $ps$. 
Spherical average over all $\vec Q$ 
with the same modulus has been performed for all $F(Q,t)$. 
}
\label{fig:fqt-T}
\end{figure}

\begin{figure}
\caption{$F(Q,t)$ at $T=207 K$. $Q$ vectors are measured in units of 
$Q_o=0.111 nm^{-1}$.   Solid lines are calculated according to
Eq.\protect\ref{eq:K}}
\label{fig:fqt-Q}
\end{figure}

\begin{figure}
\caption{Fitting parameters of $F(Q,t)$ at 
$T=207$, $T=210$, $T=215$ and $T=225 K$
according to the
stretched exponential function Eq.\protect\ref{eq:K}. Symbols as in Fig. \protect\ref{fig:fqt-T}.
}
\label{fig:fit}
\end{figure}

\begin{figure}
\caption{Fitting parameters of $F(Q,t)$ for hard-spheres 
(PY) from the numerical solution of the MCT equations,
according to the
stretched exponential function Eq.\protect\ref{eq:K}.
}
\label{fig:fit-py}
\end{figure}

\begin{figure}
\caption{$\tau_{K}(Q)/\tau_{K}(Q_{FSDP})$  
as a function of $Q$. Note that in a large $Q$ range, all curves can
be scaled on one unique curve. The continuous line is the
$Q$ dependence of collective decay times based on the
de Gennes narrowing hypothesis.
Symbols in the inset show the scaling
factor ($\tau_{K}(Q_{FSDP})$) as a function of $T$.
The full curve is the power law $\sim |T-T_c|^{2.7}$, to highlight that
the $T$ dependence of the scaling coefficient  
is compatible with MCT predictions, with the
predicted $\gamma$ value. $T_c=202 K$} 
\label{fig:tau-T}
\end{figure}

\begin{figure}
\caption{ 
Master curve for $F(Q_{FSDP},t)$ (A), $F(Q_{SQMAX},t)$ (B),
$F(Q_{SQMIN},t)$ (C). Symbols as in Fig.\protect\ref{fig:fqt-T}. 
Full line is the $beta$-correlator (the solution of the MCT equation 
\protect\ref{eq:equation} in the $\beta$-region ($g(t)$)).
The dashed line indicates the $F(Q,t)$ value chosen for
scaling the different $T$. 
The three selected $Q$ have been chosen to highlight the 
point concerning the
$Q$ dependence of the validity of the leading expansion $t^b$.
The validity of the $\beta$-correlator $g(t)$ is a priori 
limited to the region where $F(Q,t)-F_{EA}$ is small.
For clarity reason we have plotted $g(t)$ in a larger range.
}
\label{fig:master}
\end{figure}

\begin{figure}
\caption{Fit according to Eq. \protect\ref{eq:von} of the $T=207 K$
correlators limited to the early $\alpha$-relaxation region 
($7 ps~ \langle ~t~ \langle ~800 ps$ ), with $b=0.5$. 
The fitting parameters are shown in
Fig.\protect\ref{fig:h1h2}}
\label{fig:fitdue}
\end{figure}

\begin{figure}
\caption{
$Q$ dependence of the fitting parameters $h_{(1)}$, $h_{(2)}$ 
to $F(Q,t)$ at $T=207 K$. (See Eq.\protect\ref{eq:von}}).
 $h_{(1)}$ and  $h_{(2)}$ have been multiplied by $\tau^b$ and
$\tau^{2b}$ respectively to take into account the difference between
$\hat t$ and real time. The arbitrary $Q$ independent value of 
$\tau$ has been
chosen to bring  $h_{(1)}$ and  $h_{(2)}$ in the same scale as
the data of Fig.\protect\ref{fig:h1h2-py}, to make the qualitative comparison
between the two set of data easier.
\label{fig:h1h2}
\end{figure}

\begin{figure}
\caption{
$Q$ dependence of $F_{EA}$, $h_{(1)}$ and $h_{(2)}$
for Hard-Spheres (PY) at the critical packing fraction 
$\eta=0.51582 $. The PY solution for $S(Q)$ is shown for reference. R is the
hard-sphere radius.
}
\label{fig:h1h2-py}
\end{figure}

\setcounter{figure}{0}

\eject

\begin{figure}
\centerline{\psfig{figure=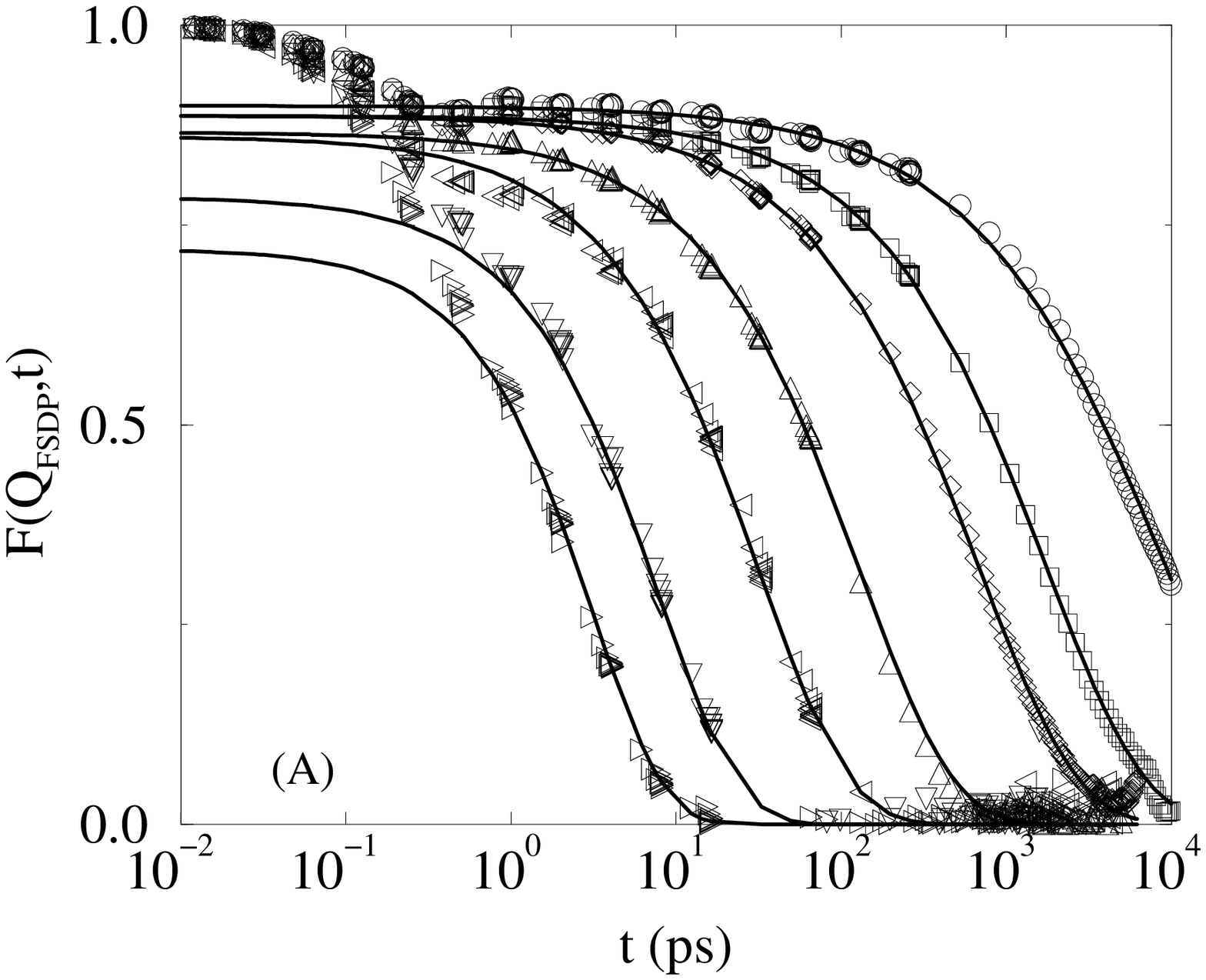,height=10cm,width=15cm,clip=,angle=0. }}
\centerline{\psfig{figure=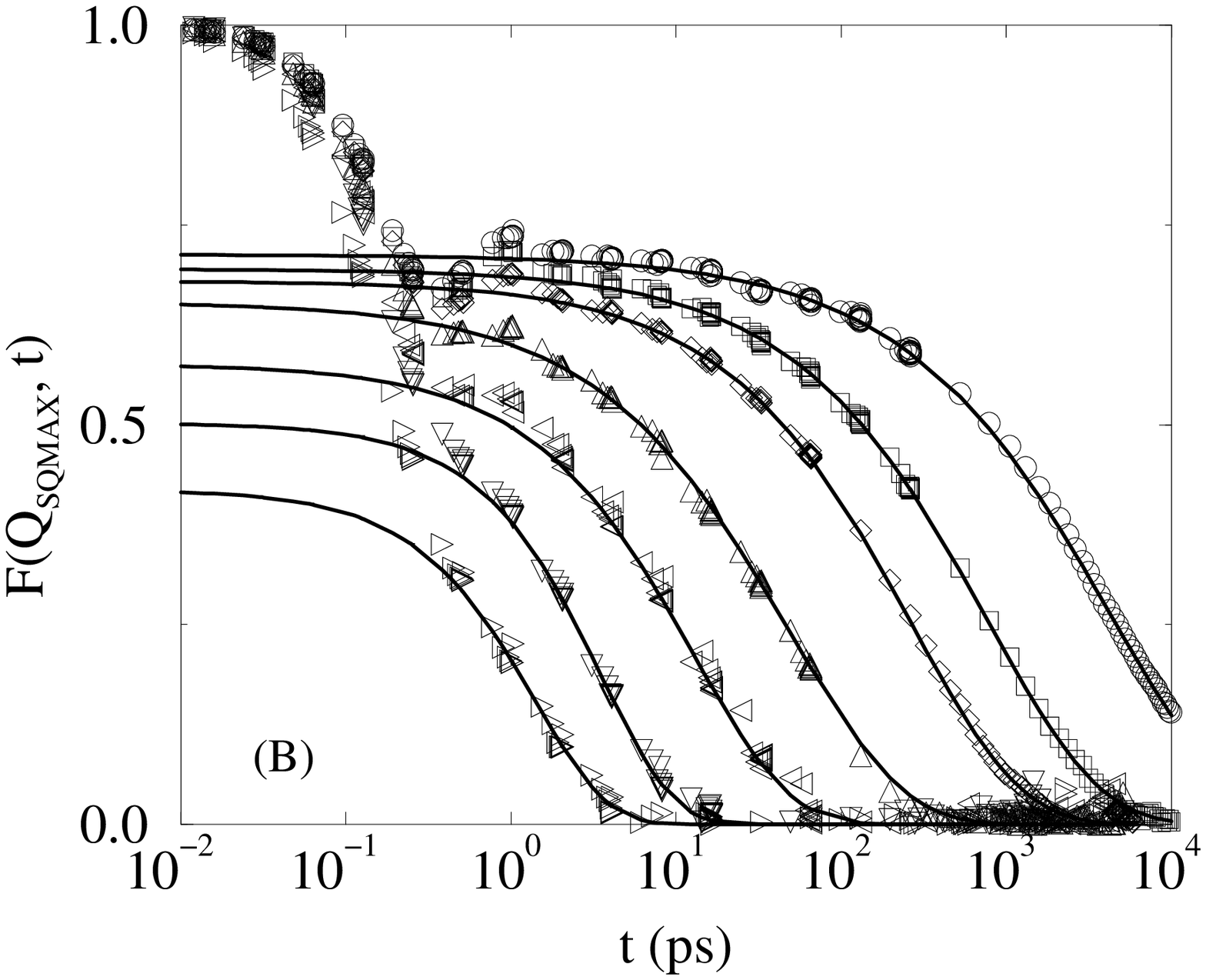,height=10cm,width=15cm,clip=,angle=0. }}
\caption{F. Sciortino et al}
\end{figure}

\eject

\begin{figure}
\centerline{\psfig{figure=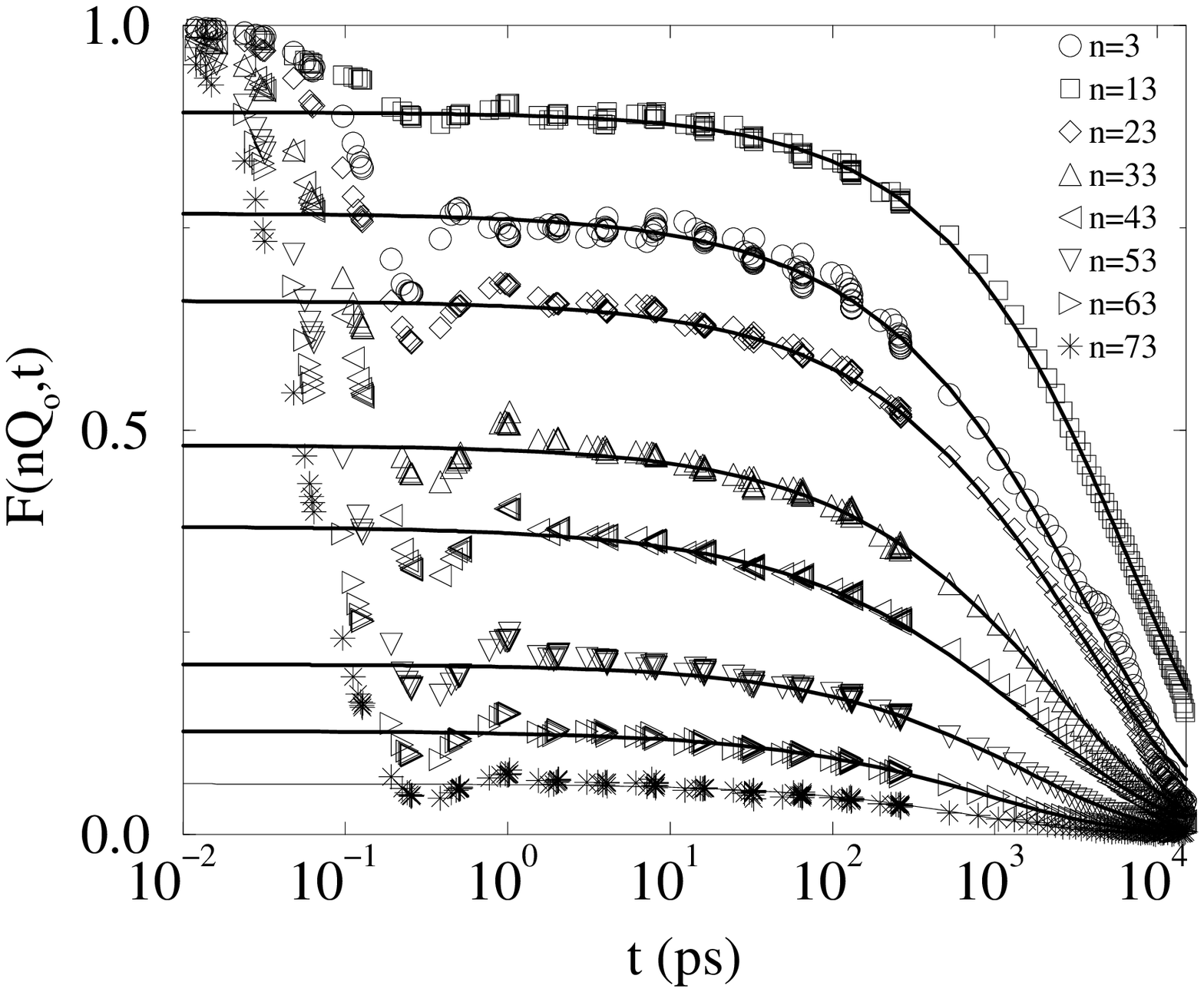,height=10cm,width=15cm,clip=,angle=0. }}
\caption{F. Sciortino et al}
\end{figure}

\eject

\begin{figure}
\centerline{\psfig{figure=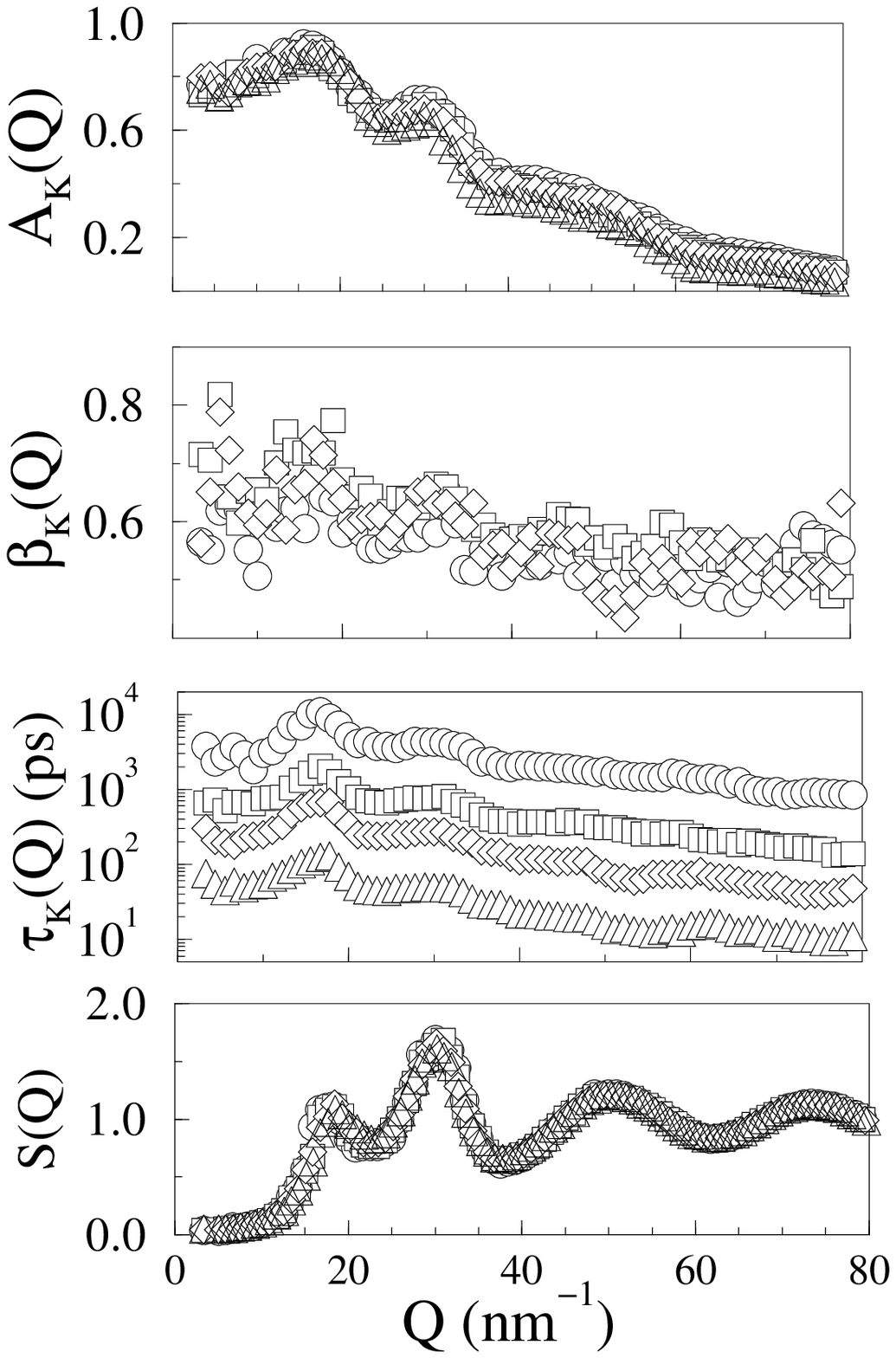,height=20cm,width=18cm,clip=,angle=0. }}
\caption{F. Sciortino et al}
\end{figure}
\eject

\begin{figure}
\centerline{\psfig{figure=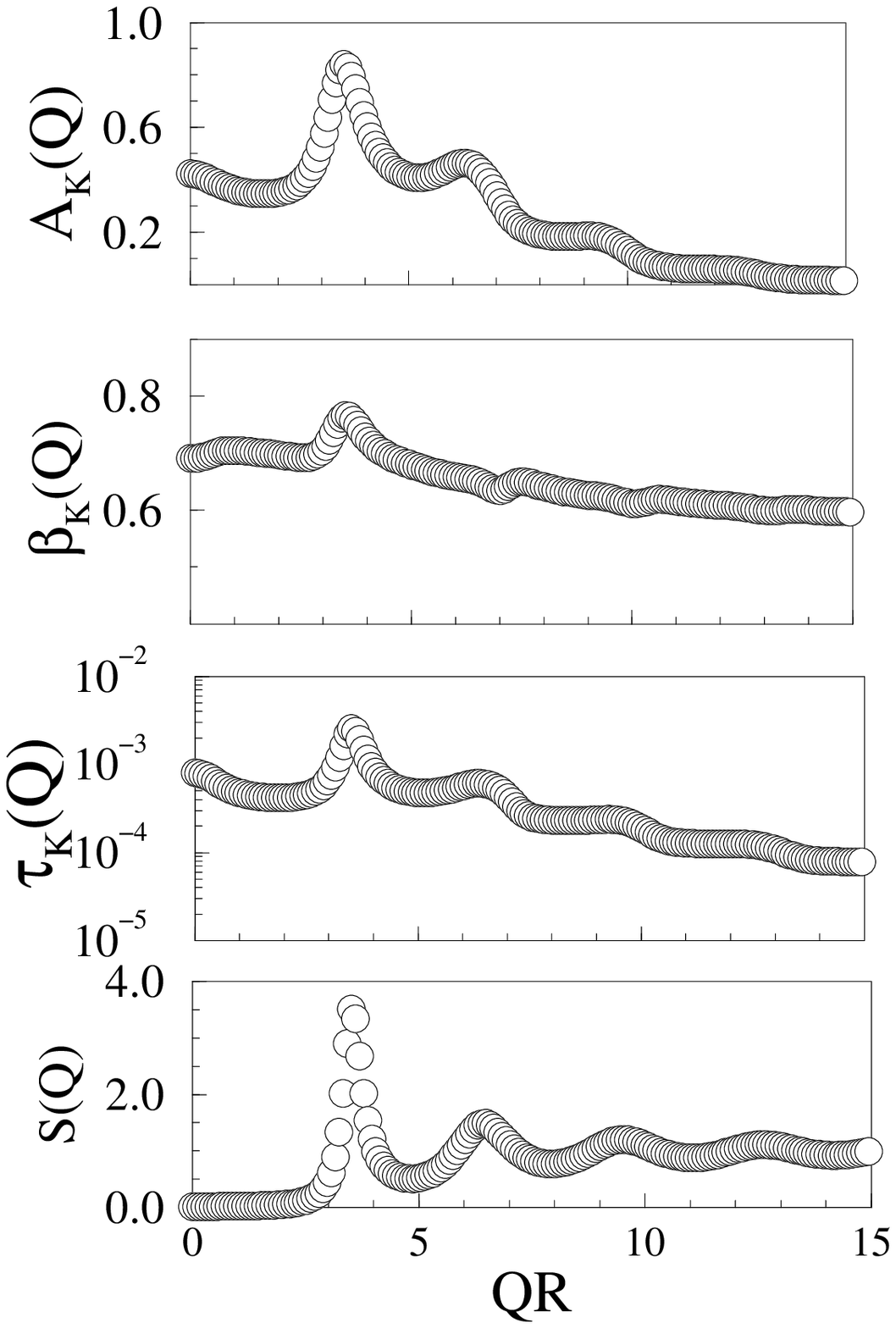,height=20cm,width=18cm,clip=,angle=0. }}
\caption{F. Sciortino et al}
\end{figure}
\eject

\begin{figure}
\centerline{\psfig{figure=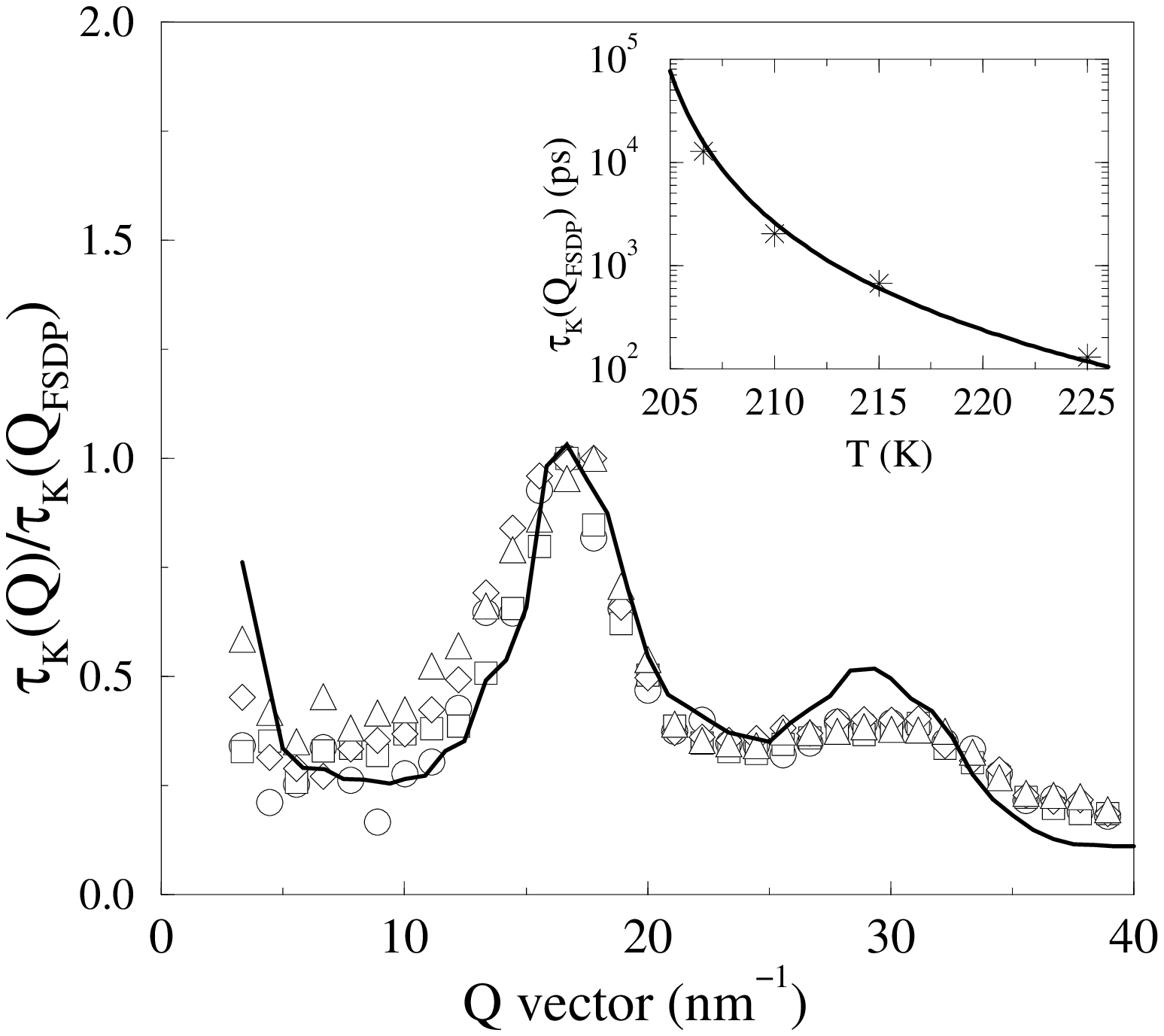,height=14cm,width=20cm,clip=,angle=0. }}
\caption{F. Sciortino et al}
\end{figure}
\eject

\begin{figure}
\centerline{\psfig{figure=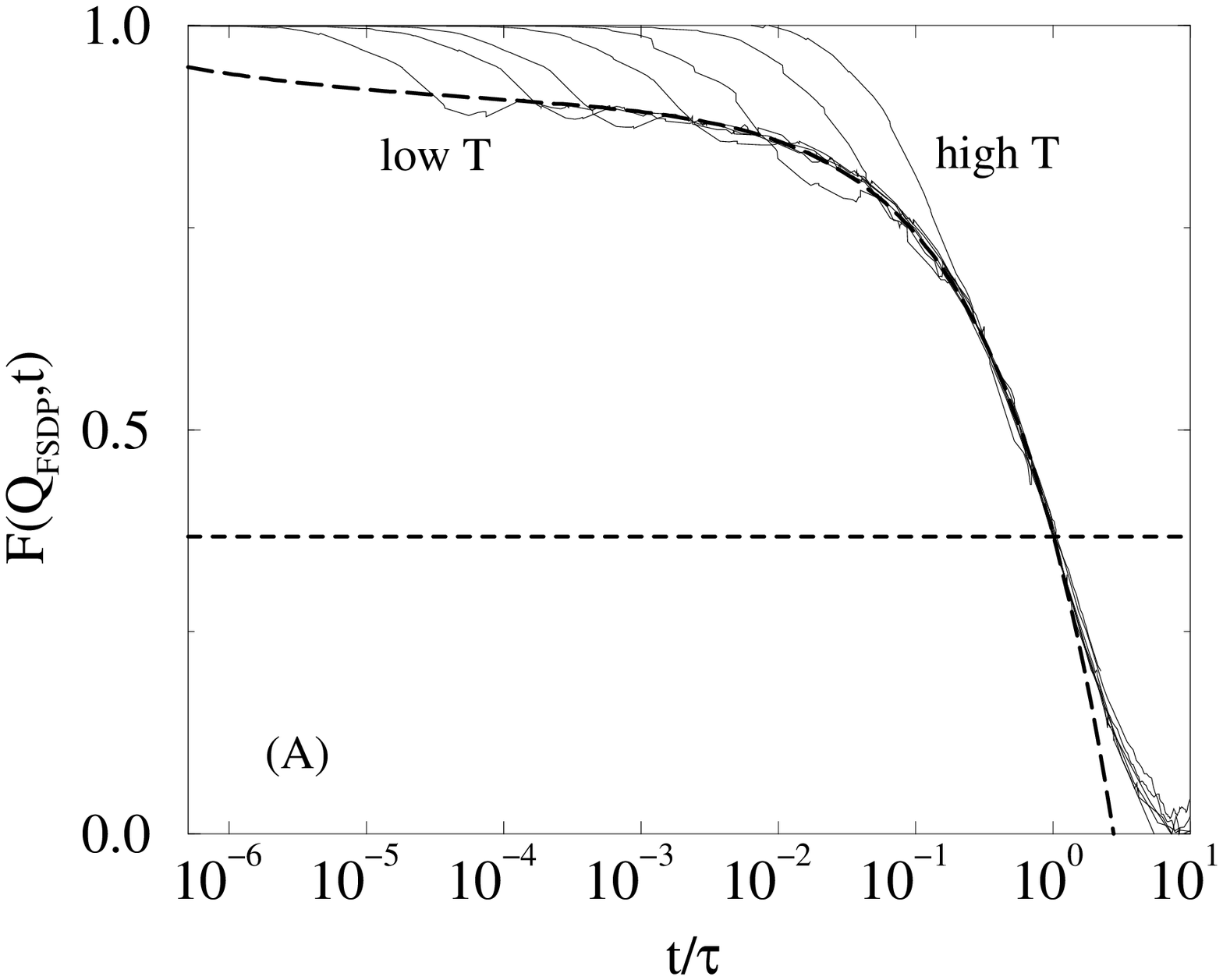,height=7cm,width=10cm,clip=,angle=0. }}
\centerline{\psfig{figure=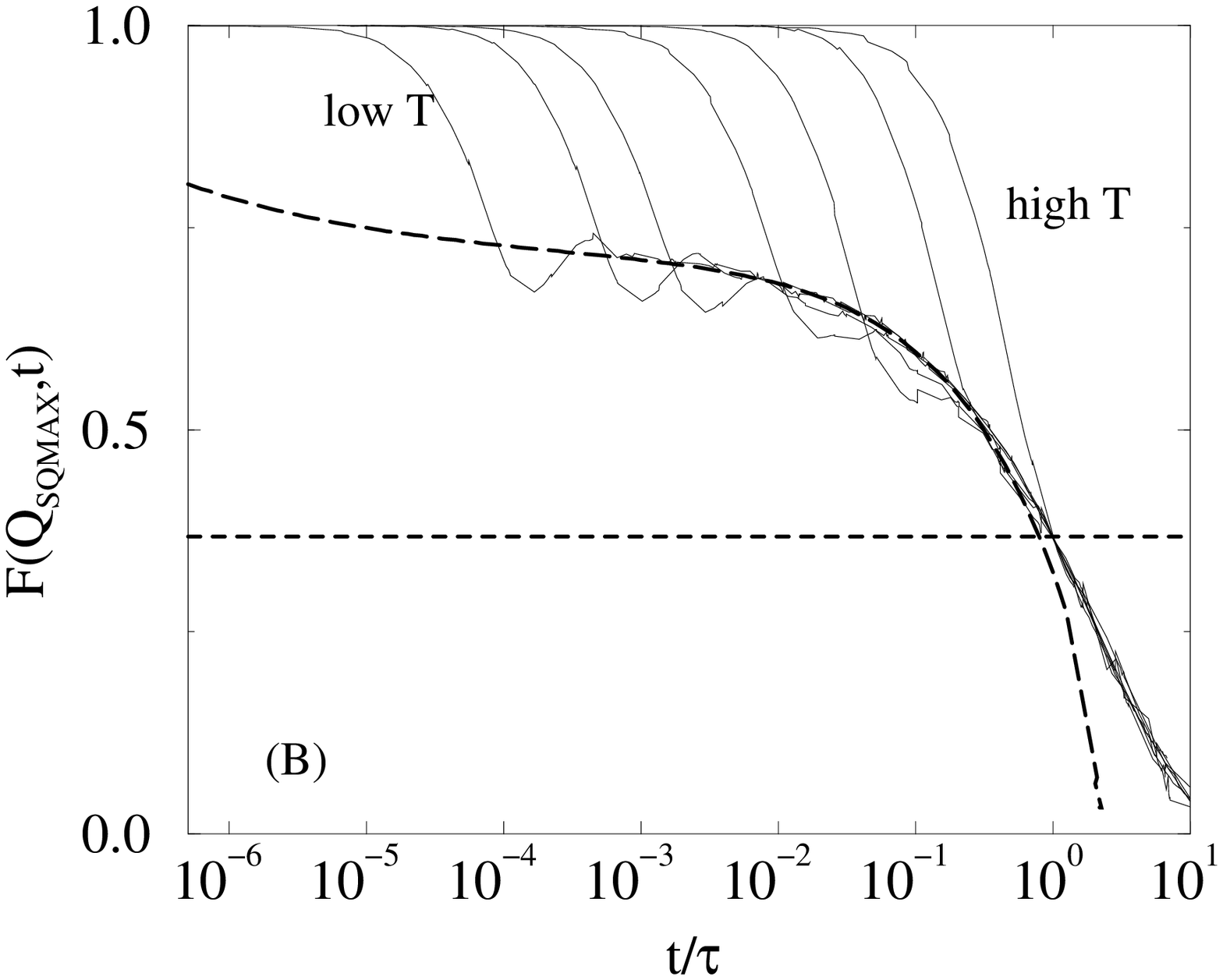,height=7cm,width=10cm,clip=,angle=0. }}
\centerline{\psfig{figure=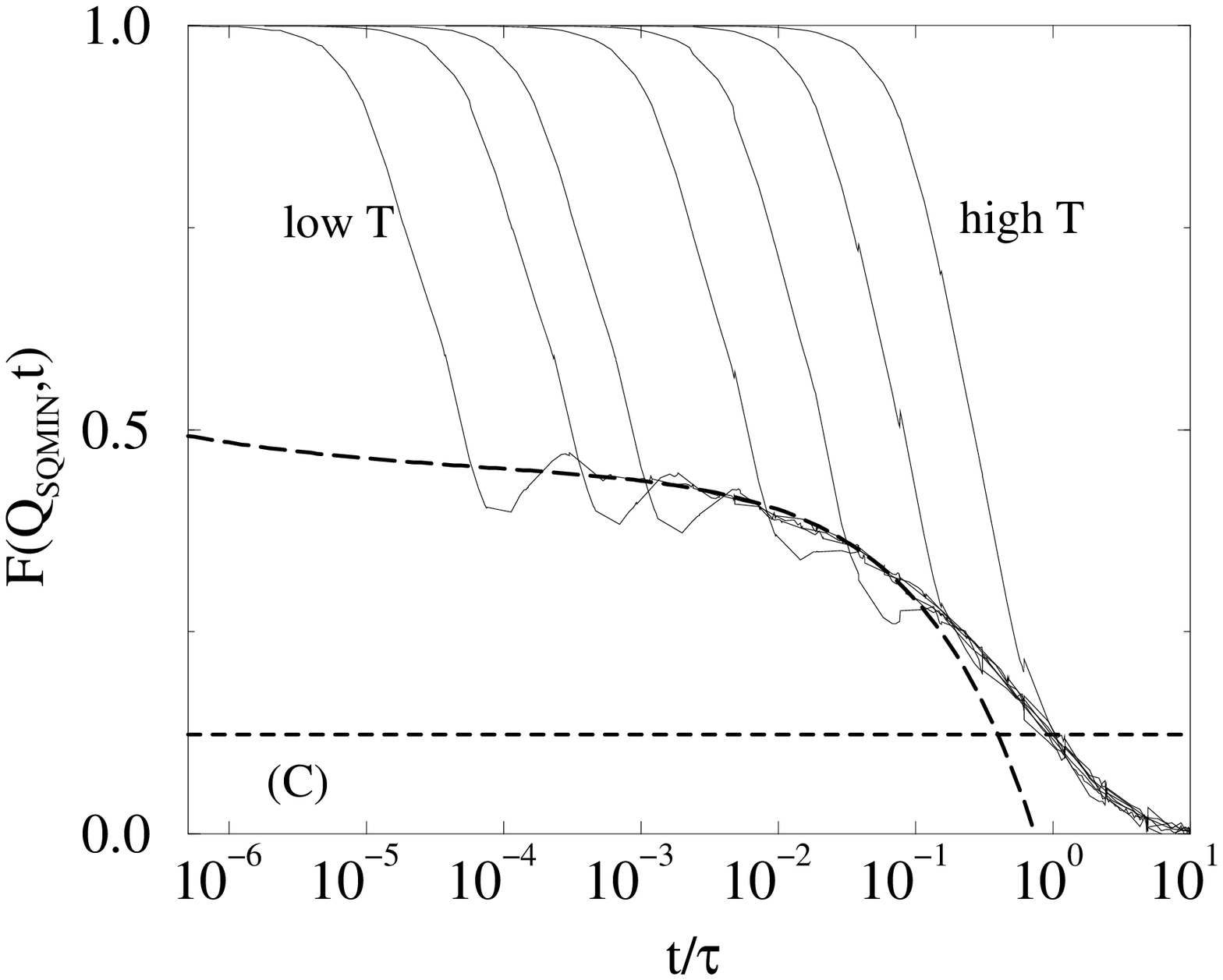,height=7cm,width=10cm,clip=,angle=0. }}
\caption{F. Sciortino et al}
\end{figure}
\eject

\begin{figure}
\centerline{\psfig{figure=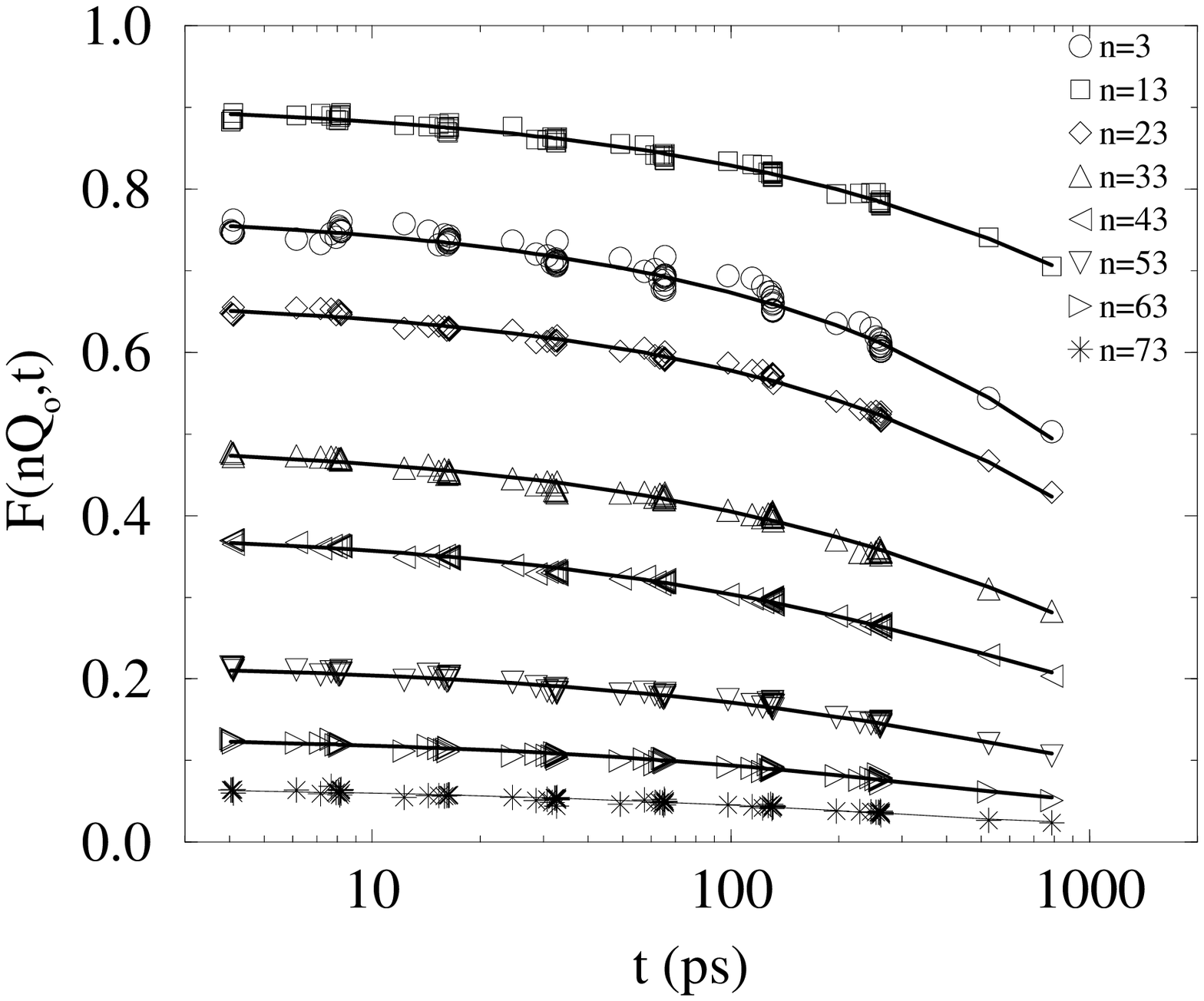,height=15cm,width=20cm,clip=,angle=0. }}
\caption{F. Sciortino et al}
\end{figure}
\eject

\begin{figure}
\centerline{\psfig{figure=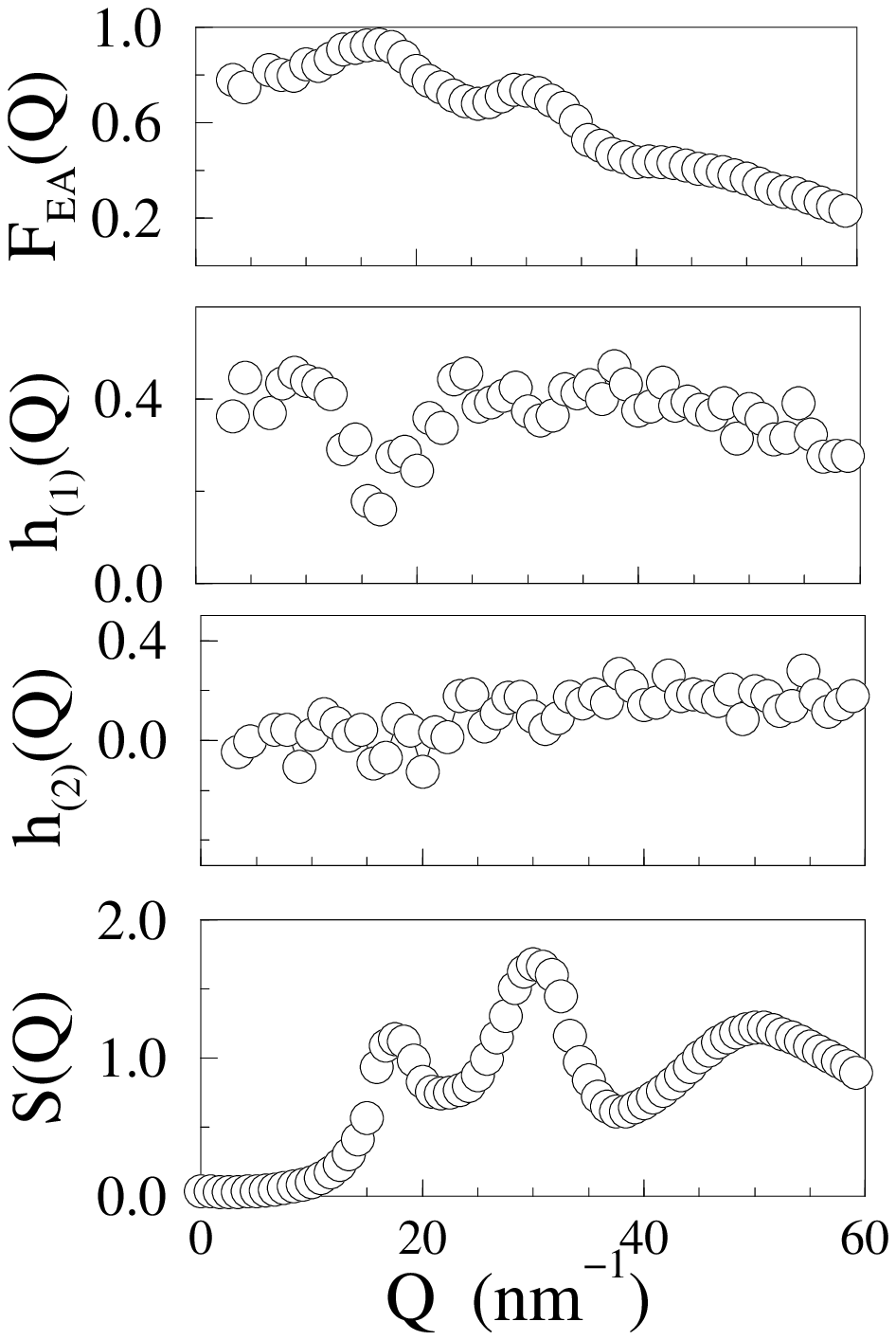,height=20cm,width=22cm,clip=,angle=0. }}
\caption{F. Sciortino et al}
\end{figure}
\eject

\begin{figure}
\centerline{\psfig{figure=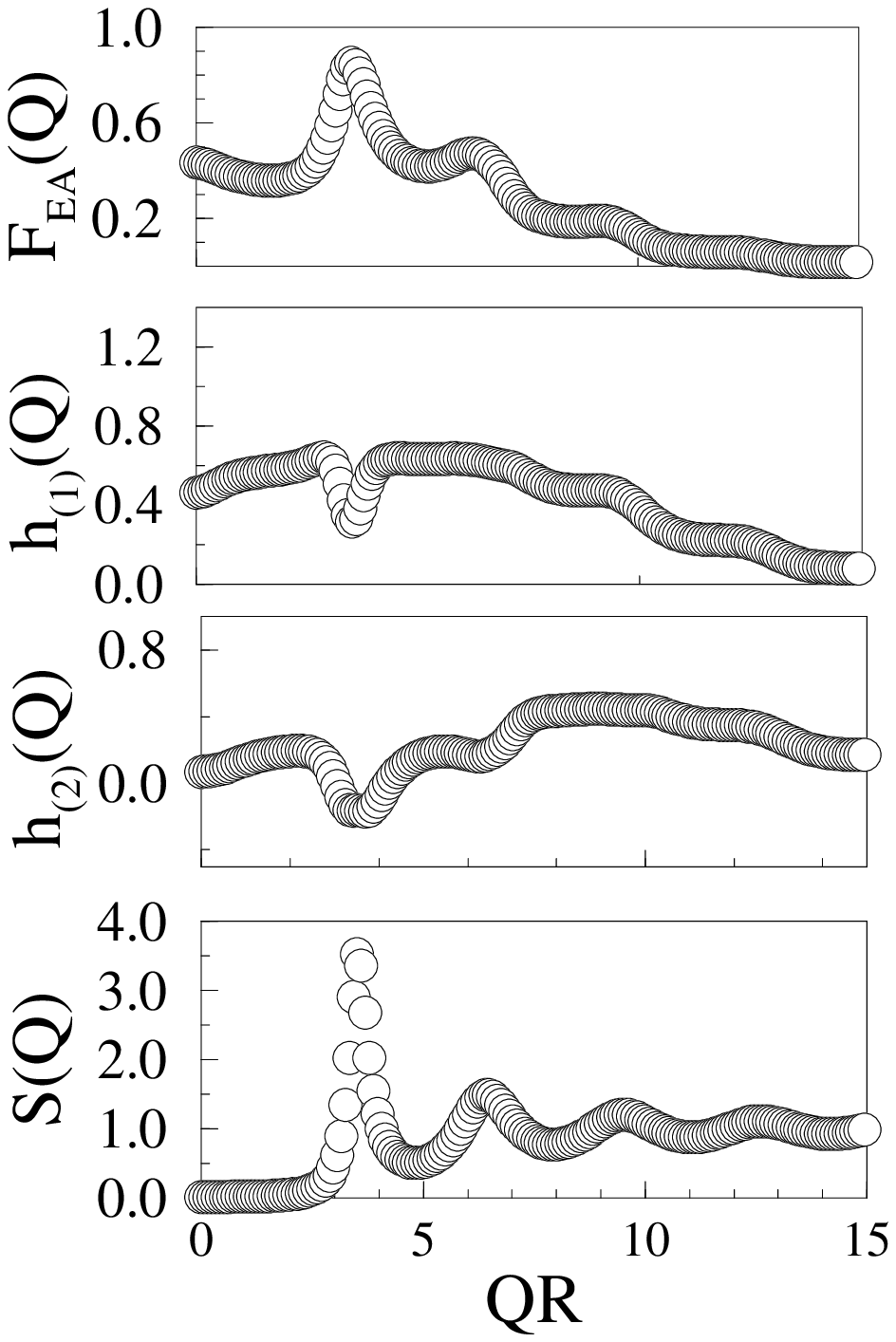,height=20cm,width=22cm,clip=,angle=0.
}}
\caption{F. Sciortino et al}
\end{figure}
\eject

\end{document}